\documentclass[submission,copyright]{eptcs}
\providecommand{\event}{FESCA 2017} 
\usepackage[T1]{fontenc}
\usepackage[utf8]{inputenc}
\usepackage{verbatim}
\usepackage{graphicx}
\usepackage{breakurl}             
\usepackage{caption}
\usepackage{subcaption}

\makeatletter

\providecommand{\tabularnewline}{\\}

\makeatother
\def\titlerunning{Model-based Testing of the Java Network API}
\def\authorrunning{C.\ Artho \& G.\ Rousset}

\begin{document}

\title{Model-based Testing of the Java network API}

\author{Cyrille Artho\institute{School of Computer Science and Communication\\KTH, Stockholm, Sweden}\institute{Information Technology Research Institute\\AIST, Osaka, Japan}
\and Guillaume Rousset\institute{University of Nantes\\Nantes, France}}
\maketitle
\begin{abstract}
Testing networked systems is challenging. The client or
server side cannot be tested by itself. We present
a solution using tool ``Modbat'' that generates test cases for 
Java's network library java.nio,
where we test both blocking and non-blocking network functions.
Our test model can dynamically simulate actions in multiple worker and client
threads, thanks to a carefully orchestrated design that covers non-determinism
while ensuring progress.

\end{abstract}

\section{Introduction}

\emph{Model-based testing} derives concrete test cases from an abstract
test model~\cite{binder2000testing,utting06}. Tools derive test sequences from a model
that specifies possible test executions and expected results~\cite{artho-hvc2013,Claessen:2000,Jacky:2007:MST:1349741,scalacheck,utting06}.
We use Modbat~\cite{artho-hvc2013} because it offers an embedded
domain-specific language~\cite{wampler2009programming} that combines
extended finite-state machines~\cite{Cheng:1993:AFT:157485.164585}
with low-level monitoring code written in Scala~\cite{Odersky2008}.

Our work on testing Java's non-blocking networking (package \texttt{java.nio})
extends previous work where we tested a custom version of that library and
found several hidden defects~\cite{artho-ase2013}. That custom library
was designed to be compatible with Java Pathfinder (JPF)~\cite{visser:03}
by working with JPF extension \texttt{net-iocache} to backtrack the
effects of network input/output~\cite{leungwattanakit2014}. It stores
the effects of network operations in memory, and replays them from previously
stored data after backtracking.

In \texttt{java.nio}, input/output actions can be blocking or
non-blocking~\cite{jdk8}.
\emph{Blocking} actions suspend the active thread
until the result is directly returned to that thread.
\emph{Non-blocking} actions return immediately, but the result may be
incomplete, requiring another call later.

Our contributions are as follows:
\begin{itemize}
\item We simulate possible concurrency on the server side (worker threads) and multiple client sessions in parallel, and ensure a proper orchestration of all activities.
\item We show how to model blocking and non-blocking functions in
Java's network library \texttt{java.nio}, where non-blocking functions
may return an incomplete result.
\end{itemize}
This paper is organized as follows: Section~\ref{sec:Background}
gives the background of this work. Section~\ref{sec:client-server}
describes our client/server model for the Java network library.
Section~\ref{sec:Conclusion-and-future-work}
concludes and outlines future work.

\section{Background\label{sec:Background}}


\subsection{Modbat}

Modbat provides an embedded domain-specific language~\cite{wampler2009programming}
based on Scala~\cite{Odersky2008} to model test executions in complex
systems succinctly~\cite{artho-hvc2013}. System behavior is described
using extended finite-state machines (EFSMs)~\cite{Cheng:1993:AFT:157485.164585}.
An EFSM is a finite-state machine that is extended with variables,
enabling functions (preconditions), and update functions (actions)
for each transition. Results of actions on the system under test (SUT)
can be checked using assertions inside the update function.

Test cases are derived by exploring available transitions, starting
from the initial state. A test case continues until a configurable
limit is hit or a property is violated. Properties include unexpected
exceptions and assertion failures. Assertions encode requirements,
typically safety properties, and are used to check the result of a
function call within a model transition.
Modbat also supports exceptions and non-deterministic outcomes:
If an exception or unexpected result
occurs during a transition, its target state can be
overridden with a different (exceptional) state~\cite{artho-hvc2013}.

Finally, Modbat offers a \texttt{launch} function, which initializes
a new child model. If multiple models are active at the same time,
they are executed using an interleaving semantics.

\subsection{The Java Network Library}

Java offers non-blocking input/output (I/O) over TCP/IP network sockets
as part of the \texttt{java.nio} package~\cite{java-api}. Two components
are essential for this work:%

\begin{enumerate}
\item \emph{Channels} represent connection entities. These include server-side
ports that can accept an incoming connection (\texttt{ServerSocketChannel})
and connection handles to send and receive data over an active connection
(\texttt{SocketChannel}).
\item \emph{Selectors} can query multiple channels at once on their availability,
chosen by using \emph{selection keys.}
\end{enumerate}
{} Blocking calls suspend the active thread until the complete result
is returned; non-blocking calls return immediately, but with a possibly
incomplete result. The application programming interface (API) of
\texttt{java.nio} allows switching between blocking and non-blocking
modes at any time.

\subsection{Related Work}

Unit testing experienced a widespread rise in software development
in the late 1990s~\cite{junit}. While unit testing automates
test execution, model-based testing automates test design~\cite{binder2000testing,utting06}.
Instead of designing individual test cases, test models describe
entire sets of possible tests. More test tools than can be described here
exist, based on state
machines~\cite{artho-hvc2013,Jacky:2007:MST:1349741,utting06} or constraint
specifications~\cite{Claessen:2000,scalacheck}.
Test models (as well as unit tests)
are usually designed based on the specification~\cite{utting06}.

For the Java API, related work~\cite{luo2014rv} presents a systematic
formulation of the specification of the Java API, and a run-time monitor
implementation which monitors the correct usage of the API. While our
work implements a monitor which verifies the correctness of
implementation of the API itself (rather than its usage by an application),
systematic ways to construct the
specification of the API greatly interest us.

\section{Test Model for the Java Network API\label{sec:client-server}}

We start with the overall organization of our test model and add
selector-based, non-blocking I/O later.

\subsection{Minimalist Model for Client/Server Connections}

A server using blocking I/O uses multiple threads to handle multiple
connections at the same time: A server \emph{main} thread accepts
incoming requests and then spawns a \emph{worker} thread, which uses
a given connection handle and deals with the request. Each client
is typically an independent process that first connects to the server
and then communicates with it. For the purpose of testing, we
launch client models as child instances, which simulate external processes
while being executed inside the test harness.

\begin{figure}
\begin{centering}
\begin{tabular}{cr}
\begin{minipage}[c]{0.48\columnwidth}%
\includegraphics[scale=0.9]{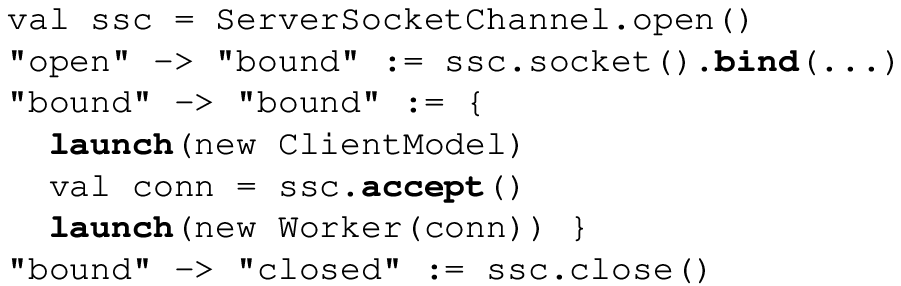}
\end{minipage} & %
\begin{minipage}[c]{0.48\columnwidth}%
\begin{flushright}
\includegraphics[scale=0.5]{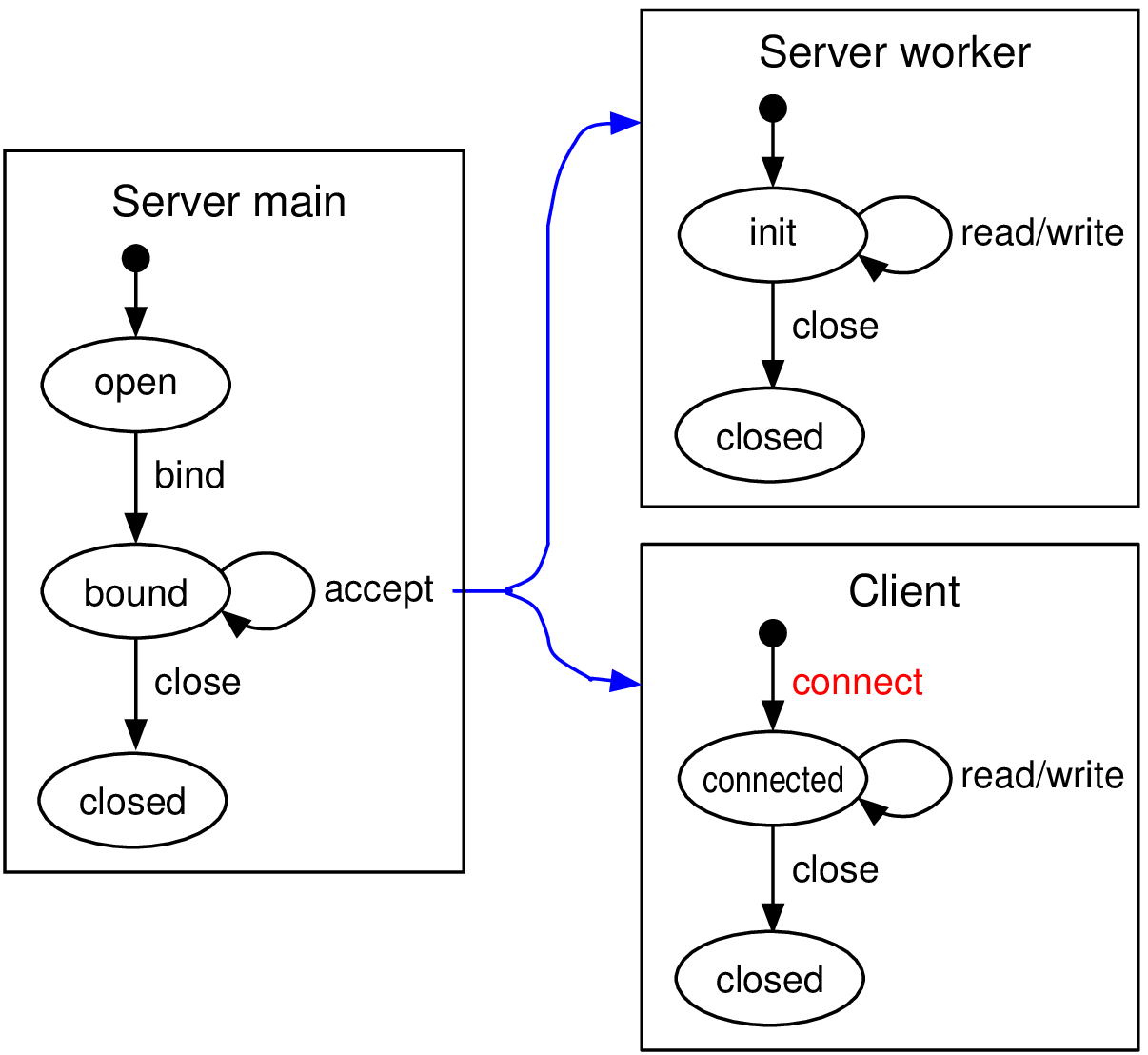}
\par\end{flushright}%
\end{minipage}\tabularnewline
\end{tabular}
\par\end{centering}
\caption{Minimalist model for client/server connections, with the server
main model code (left) and a visualization of all three components
(right). The second transition first launches a client model, then
calls ``accept'' on the \texttt{ServerSocketChannel}, and finally
passes the handle for the incoming connection to a newly launched server worker
model.\label{fig:cl-serv-model}}
\end{figure}

Figure~\ref{fig:cl-serv-model} shows
this parallel composition of dynamically instantiated models. It mirrors
a server that accepts incoming client connections and then delegates
the connection handle to a worker thread.
The crucial part
is transition ``bound'' $\rightarrow$ ``bound''. To ensure a
correct handshake that establishes a working connection, a client
model that connects back to the server in its constructor has to be
launched first. The call to \texttt{connect} is executed in the constructor
of the model; this ensures that the subsequent call to \texttt{accept}
succeeds (because the operating system queues the pending client connection
request). After that, the connection handle is passed to a new server
worker model instance, which uses that connection.

\subsection{Server Implementation Using Selectors}

High-performance servers use non-blocking, selector-based I/O, to
handle many connections in a single execution thread~\cite{c10k}. 
The server usually calls \texttt{select}
in an infinite loop,
which returns a set of available channels from which data can
be consumed~\cite{jdk8,java-nio-tutorial}.
Available operations include \texttt{accept}, to handle
a new incoming connection, and \texttt{read}, to handle new data on
an existing connection. The complexity of concurrency in the previous
architecture is replaced by non-determinism w.\,r.\,t.\ the outcome
of each \texttt{select} call and possibly incomplete \texttt{read}
and \texttt{write} operations that require careful buffer management.

\subsection{Server Main and Worker Models Using Selectors}

Our Modbat model for \texttt{java.nio} expands on an earlier
version, which was used to uncover defects in a custom version of the
\texttt{java.nio}
library but did not use multiple connections~\cite{artho-ase2013}.

\begin{figure*}
\begin{centering}
\includegraphics[angle=-90,scale=0.53]{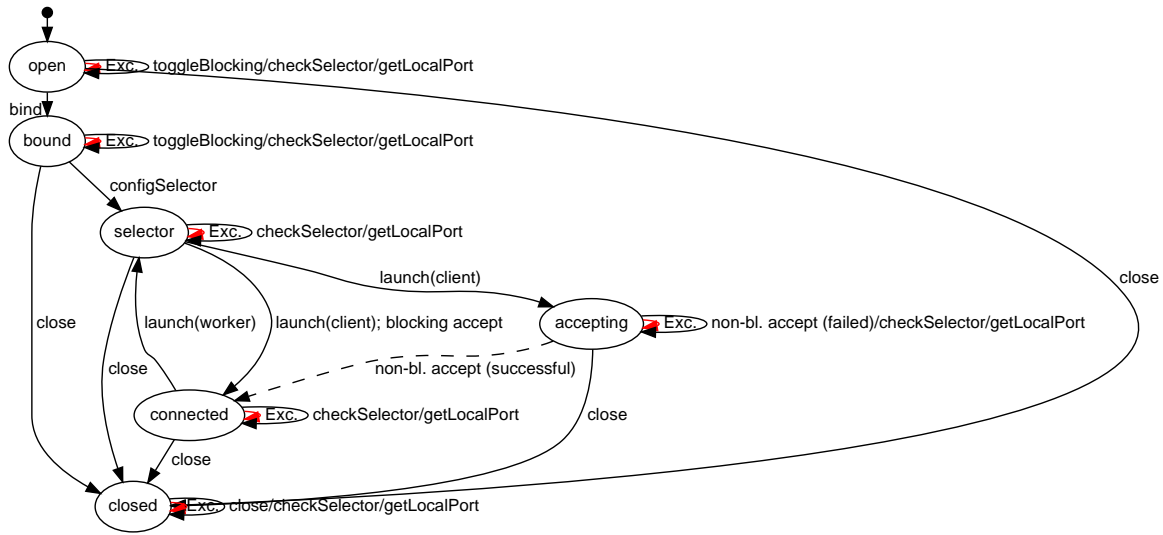}
\par\end{centering}
\caption{Detailed model of the server main thread.
Labeled nodes are states, black solid arrows
are model transitions. The dashed transition represents
a successful non-blocking accept call, which is modeled as a non-deterministic outcome.
Red arrows represent exceptions
for operations that are disallowed in some states.
`/'' shows alternatives for a transition, ``;'' a concatenation of two actions.
\label{fig:java-nio-main}}
\end{figure*}

The detailed server main model in Figure~\ref{fig:java-nio-main}
refines the original simple model with selectors and non-blocking
calls, which have to be repeated until they succeed.
It starts with a fully initialized instance of \texttt{ServerSocketChannel}.
Each state has several self-transitions, which do not affect the state
of the model or the corresponding object that is tested.
Self-transitions may change between blocking and
non-blocking mode (toggleBlocking), check the state of the selector
(checkSelector) or the local port number (getLocalPort), or perform
an operation that is not allowed in a given state and that results
in an exception (red transitions labeled ``Exc.'').
A ``successful'' path through the test model
first binds the channel to an address and network port, then
configures the selector, and then accepts one or more client connections
before shutting down the service by calling \texttt{close} on the
\texttt{ServerSocketChannel}. Calls to \texttt{accept} in non-blocking
mode may initially fail because the client model may not be ready when
the call is made. In that case, the model stays in an intermediate
``accepting'' state. In that state, the call to \texttt{accept}
can be repeated until the
operation is successful. As the outcome depends on network latency,
a successful result is modeled as a non-deterministic outcome in Modbat,
shown as a dotted transition in Figure~\ref{fig:java-nio-main}.
The model proceeds to state ``connected'' if the returned connection
object is initialized (non-null) and stays in state ``accepting'' otherwise.

\begin{figure*}
\begin{centering}
\begin{subfigure}[b]{0.75\textwidth}
\centering
\includegraphics[angle=-90,scale=0.55]{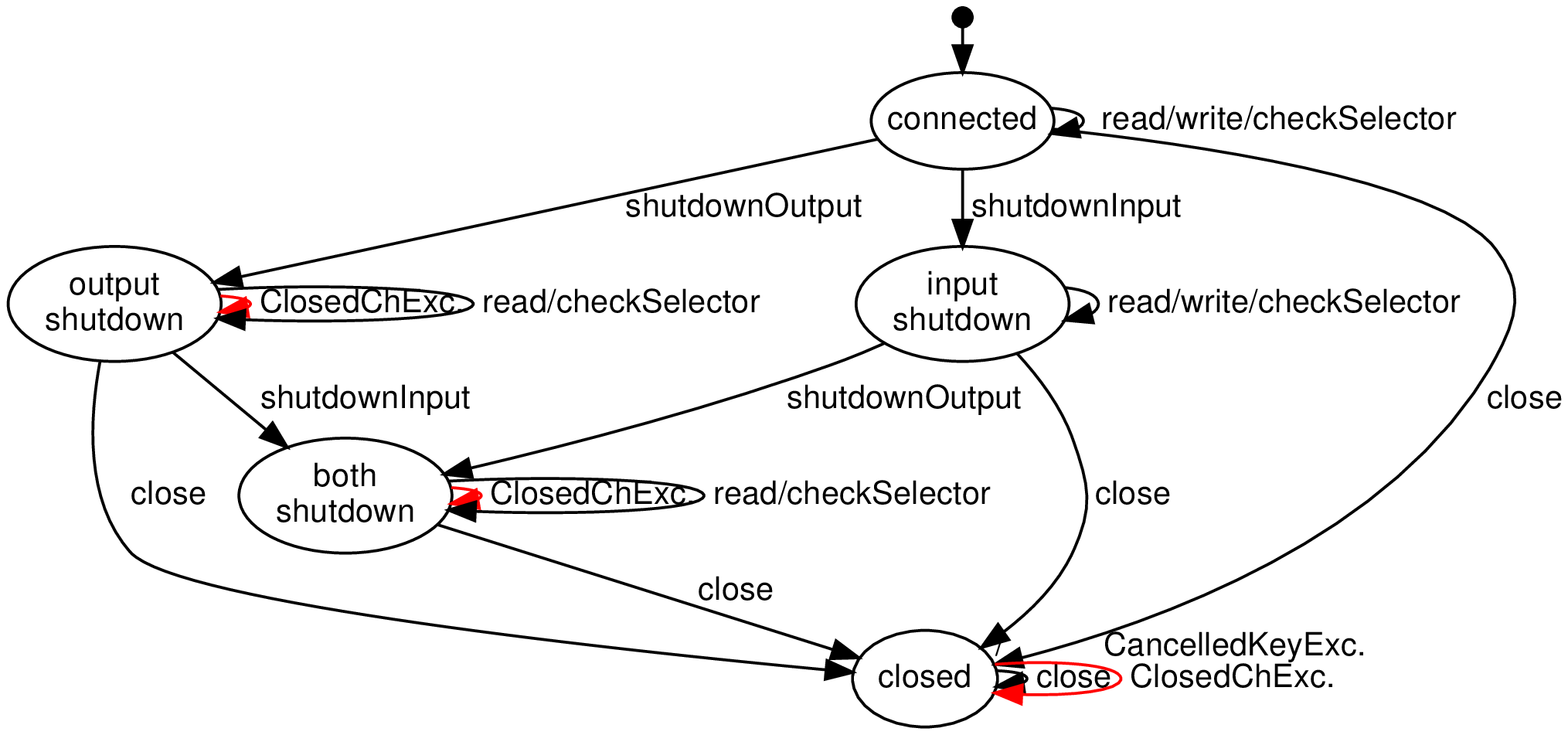}
\caption{Server-side connection model}\label{fig:serv-conn}
\end{subfigure}\hfill
\begin{subfigure}[b]{0.25\textwidth}
\centering
\includegraphics[angle=-90,scale=0.55]{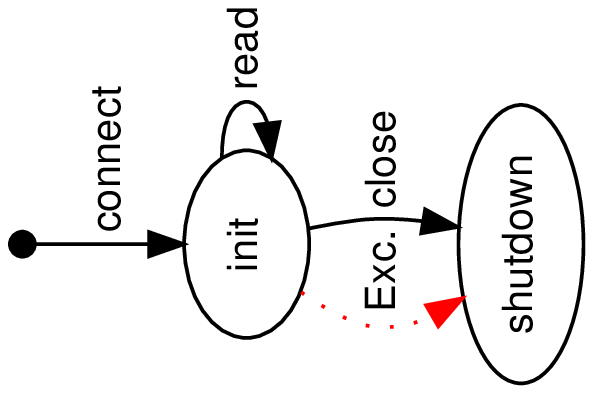}
\vspace{2\baselineskip}
\caption{Client model}\label{fig:client-conn}
\end{subfigure}
\par\end{centering}
\caption{Test models representing a server connection
(``worker'', left) and a client connection (right). 
Selector calls are possible as long as the channel is not closed.
The client model reads until either an exception occurs, or by a non-deterministic
model choice closes the connection.\label{fig:java-nio-session}}
\end{figure*}

Because selector and read/write calls can be made on each connection
independently, a separate ``worker'' model simulates such calls (see
Figure~\ref{fig:serv-conn}). In this model, the generated model
instances simulate interleaved selector usages.
Each connection model starts with an initialized connection, from which
at can proceed by shutting down the input or output channel, or both
by closing the connection. In each state, self-transitions can read
or write on the channel, or check the state of its selectors.
Once a channel is partially or fully closed, several operations result
in an exception.

Each server-side connection model instance is paired with
a minimalist client instance (see Figure~\ref{fig:client-conn}).
This ensures a connection to the server socket is eventually established,
which is necessary for testing networked software
components, as they cannot be executed in isolation.

We check that the return values and exceptions thrown are consistent
with the model-side view of the overall system. Due to network latency,
it may be possible that data is not available in cases where an ideal
network could provide data; our assertions take this into account.
For example, the number of bytes read may be less or equal than the
number of bytes available in the channel.
The opposite case, a return value suggesting the availability of data
where no data is believed (by the model) to exist, results in a property
violation that is shown as an error trace by Modbat~\cite{artho-ase2013}.

\subsection{Challenges}

Side effects of test actions
pose a challenge. If a test does not clean up all resources
when it ends, dependencies between tests may arise.
We observed this in cases where a test could not be replayed in isolation
but only if it was part of a larger sequence of multiple tests.
We eventually arrived at what we think is a correct
model by executing half a million test cases against the system, without
false positives.

Test cases for the Java network API execute quickly, but ephemeral
ports that handle a connection are exhausted after a while. On today's
systems, a few tens of thousands of such ports are available; their
number can be increased only slightly with a custom kernel configuration.
A lack of available ephemeral ports results in a slowdown of test
execution (until closed ports are made available again).

In the model described in this paper, the behavior of each connection is independent of other
connections, so the test oracle is entirely local to each model instance,
and independent of interleavings of messages sent over the network.
If this is not the case, the oracle has to consider all possible interleavings
of events, as described in other work~\cite{artho-icst2017}.

\section{Conclusion and Future Work\label{sec:Conclusion-and-future-work}}


We present a test model for a network API that uses
multiple parallel model instances to simulate concurrent requests.
The critical issue is to ensure progress after launching a new model instance,
or when waiting for a request. To achieve this, (1) the server side must
be ready to receive requests; (2) a client session is initiated; and
(3), in the case of the Java network API, the necessary server call
to accept the client session must be executed. Changing the order
of these steps leads to a deadlock in test execution.

Our model faithfully reflects all key operations of the Java network
library~\cite{jdk8}, and can be used to analyze different implementations
in depth; previous work reports on defects found with our approach~\cite{artho-ase2013}.

The models we present can be adapted to other client/server systems.
In the future, we want to explore more network APIs, and other uses
of Modbat as an event simulator.

Modbat and example models are available at
\url{http://people.kth.se/~artho/modbat/}.

\clearpage

\bibliographystyle{eptcs}
\bibliography{biblio}

\end{document}